\documentclass[a4paper]{article}
\usepackage[margin=2.5cm]{geometry}
\usepackage{amsmath,amssymb,amsthm}
\usepackage{algorithm,algpseudocode}
\usepackage{graphicx}
\usepackage{hyperref}
\usepackage{booktabs}
\usepackage{enumitem}
\usepackage{xcolor}
\usepackage[sorting=nyt, style=numeric, backend=biber, clearlang=true, url=false, doi=false, isbn=false, giveninits=true]{biblatex}
\title{Self-Scaled Broyden Family of Quasi-Newton Methods in JAX}
\author{%
  Ivan Bioli\textsuperscript{a,b} \and Mikel Mendibe
  Abarrategi\textsuperscript{c,d}%
  } \date{
    \footnotesize
    \textit{ \textsuperscript{a} Dipartimento di Matematica, Università degli
    Studi di Pavia, Via A. Ferrata 5, 27100 Pavia, Italy\\
    \textsuperscript{b} Dipartimento di Ingegneria Civile e Architettura,
    Università degli Studi di Pavia, Via A. Ferrata 3, 27100 Pavia, Italy\\
    \textsuperscript{c} University of the Basque Country, 48013 Bilbao, Spain\\
    \textsuperscript{d} TECNALIA, Basque Research \& Technology Alliance (BRTA),
    48160 Derio, Spain\\
    } }

\begin{document}
\maketitle

\begin{abstract}
We present a JAX implementation of the Self-Scaled Broyden family of quasi-Newton methods, fully compatible with JAX and building on the Optimistix~\cite{rader_optimistix_2024} optimisation library. The implementation includes BFGS, DFP, Broyden and their Self-Scaled variants (SSBFGS, SSDFP, SSBroyden), as well as L-BFGS, together with a Zoom line search satisfying the strong Wolfe conditions. This is a short technical note, not a research paper, as it does not claim any novel contribution; its purpose is to document the implementation and ease the adoption of these optimisers within the JAX community. The code is available at \url{https://github.com/IvanBioli/ssbroyden_optimistix} and on PyPI (\texttt{pip install ssbroyden-optimistix}).
\end{abstract}

\section{Introduction}

Optimistix~\cite{rader_optimistix_2024} is a JAX library for nonlinear solvers
providing modular and composable optimisation algorithms. While Optimistix
includes standard BFGS, DFP and L-BFGS implementations paired with a
backtracking Armijo line search, it lacks both the Zoom line search, which satisfies the strong Wolfe
conditions, and the broader family of Self-Scaled Broyden
methods \cite{al-baali_numerical_1998,
    oren_self-scaling_1974-1,kiyani_optimizing_2025}. This work provides a pure-JAX implementation of these
methods, designed to be fully compatible with the Optimistix solver interface:
the new solvers can be used as drop-in replacements, composed with existing
Optimistix descents and searches, and benefit from all JAX transformations.
Specifically, our implementation addresses the following gaps in the current
Optimistix offerings:

\begin{enumerate}[leftmargin=*]
    \item \textbf{Zoom line search.} We integrate the Zoom line search
          (Algorithm~3.6 in~\cite{wright_numerical_2006}) into Optimistix,
          ensuring that the strong Wolfe conditions are satisfied at each step.
          The implementation is adapted from
          \texttt{bagibence/zoom\_linesearch}\footnote{\url{https://github.com/bagibence/optimistix/tree/zoom_linesearch}}
          with minor modifications to fit the new Optimistix interface.

    \item \textbf{Self-Scaled Broyden family.} We implement the full Self-Scaled
          Broyden family of quasi-Newton Hessian updates, encompassing Broyden,
          BFGS, DFP, and their Self-Scaled variants SSBroyden, SSBFGS, and
          SSDFP.

    \item \textbf{Iteration counting.} We provide a driver,
          \texttt{ssbroyden\_optimistix.contrib.run}, that distinguishes
          between actual quasi-Newton iterations and internal line search steps,
          which Optimistix does not separate by design choice. This allows for
          more refined comparisons between solvers.
\end{enumerate}

\section{The Self-Scaled Broyden Family}

The Self-Scaled Broyden family of quasi-Newton methods generalises the classic
Broyden, BFGS, and DFP updates to minimize a function
$f:\mathbb{R}^N \to \mathbb{R}$ \cite{al-baali_numerical_1998,
    oren_self-scaling_1974-1,kiyani_optimizing_2025}. At each iteration $k$, these methods maintain
an approximation $H_k$ of the inverse Hessian of $f$ and compute a search
direction $d_k = -H_k \nabla f(x_k)$. After a line search determines a step size
$\alpha_k$, the iterate is updated as $x_{k+1} = x_k + \alpha_k d_k$. The
inverse Hessian approximation is then updated using the step $s_k = x_{k+1} -
    x_k$ and the gradient difference $y_k = \nabla f(x_{k+1}) - \nabla f(x_k)$.

The Self-Scaled Broyden family update is parameterised by two scalars,
$\theta_k$ and $\tau_k$. In its most general form, the update is given by
\begin{equation}\label{eq:update}
    H_{k+1} = \frac{1}{\tau_k} \left(
    H_k
    - \frac{H_k y_k y_k^\top H_k}{y_k^\top H_k y_k}
    + \phi_k \,(y_k^\top H_k y_k)\, v_k v_k^\top
    \right)
    + \rho_k\, s_k s_k^\top,
\end{equation}
where
\begin{equation*}
    \begin{split}
        \rho_k & = \frac{1}{y_k^\top s_k},                                      \\
        v_k    & = \frac{s_k}{y_k^\top s_k} - \frac{H_k y_k}{y_k^\top H_k y_k}, \\
        \phi_k & = \frac{1-\theta_k}{1 + (h_k b_k - 1)\theta_k},                \\
        b_k    & = \frac{s_k^\top B_k s_k}{y_k^\top s_k},                       \\
        h_k    & = \frac{y_k^\top H_k y_k}{y_k^\top s_k},
    \end{split}
\end{equation*}
and $B_k = H_k^{-1}$ is the corresponding approximation of the Hessian.
The parameter $\theta_k$ interpolates between the BFGS ($\theta_k=0$) and DFP
($\theta_k=1$) updates, with the more general Broyden family obtained by
computing $\theta_k$ dynamically at each iteration as:
\begin{equation*}
    \theta_k = \max \left(\theta_k^{-},\,
    \min \left(\theta_k^{+},\,\frac{1-b_k}{b_k}\right)\right),
\end{equation*}
where, letting $a_k = b_k h_k - 1$ and $c_k =
    \left(\frac{a_k}{1+a_k}\right)^{1/2}$,
\begin{equation*}
    \rho_k^{-} = \min \bigl(1,\, h_k(1-c_k)\bigr),\qquad
    \theta_k^{-} = \frac{\rho_k^{-}-1}{a_k},\qquad
    \theta_k^{+} = \frac{1}{\rho_k^{-}}.
\end{equation*}
The parameter $\tau_k$ controls the Self-Scaled variant ($\tau_k=1$ means no
scaling) and is computed as
\begin{equation*}
    \tau_k =
    \begin{cases}
        \min \bigl(\rho_k^{+}\,\sigma_k^{(1-N)},\,\sigma_k\bigr),
         & \text{if } \theta_k \le 0, \\[4pt]
        \rho_k^{+}\,\min \bigl(\sigma_k^{(1-N)},\,
        \tfrac{1}{\theta_k}\bigr),
         & \text{otherwise,}
    \end{cases}
\end{equation*}
where
\begin{equation*}
    \rho_k^{+} = \min \left(1,\,\frac{1}{b_k}\right),\qquad
    \sigma_k = 1 + \theta_k\, a_k,\qquad
    \sigma_k^{(1-N)} = |\sigma_k|^{\frac{1}{1-N}}.
\end{equation*}
By default (option \texttt{initial\_scaling=True}), the first update instead
uses the initial scaling $\tau_0 = h_0/(1 + a_0\theta_0)$. Since $B_k$ is not
available when approximating the inverse Hessian, $b_k$ is computed from the
identity $B_k s_k = -\alpha_k \nabla f(x_k)$ as $b_k = -\alpha_k\, s_k^\top
\nabla f(x_k)/(y_k^\top s_k)$, where by default (option
\texttt{alpha="geometric"}) the step length is recovered from the step as
$\alpha_k = s_k^\top d_k / (d_k^\top d_k)$, which is exact for any line search.
Table~\ref{tab:solvers} summarises the six concrete solvers obtained by choosing
$\theta_k$ and~$\tau_k$.

\begin{table}[ht]
    \centering
    \caption{Solvers implemented as special cases of the Self-Scaled Broyden family.}
    \label{tab:solvers}
    \begin{tabular}{lccl}
        \toprule
        Solver    & $\theta_k$ & $\tau_k$ & Description                 \\
        \midrule
        BFGS      & $0$        & $1$      & Classic BFGS                \\
        SSBFGS    & $0$        & computed & Self-Scaled BFGS            \\
        DFP       & $1$        & $1$      & Classic DFP                 \\
        SSDFP     & $1$        & computed & Self-Scaled DFP             \\
        Broyden   & computed   & $1$      & Broyden family (no scaling) \\
        SSBroyden & computed   & computed & Self-Scaled Broyden family  \\
        \bottomrule
    \end{tabular}
\end{table}

\subsection{Software Design}

The implementation follows a class hierarchy that mirrors the mathematical
structure of the update family, building on an \texttt{AbstractQuasiNewton}
base class adapted from Optimistix:

\begin{enumerate}[leftmargin=*]
    \item \texttt{AbstractSSBroydenFamily} implements the shared logic: Hessian
          initialisation, computation of the auxiliary quantities, the general
          update term~\eqref{eq:update}, and the choices of $\theta_k$ and
          $\tau_k$ described above. It exposes two hooks:
          \texttt{\_compute\_thetak} and \texttt{\_compute\_tauk}, which
          subclasses override to fix or compute $\theta_k$ and $\tau_k$.

    \item \texttt{AbstractSSBroyden} keeps these choices, computing both
          $\theta_k$ and $\tau_k$ dynamically. \texttt{AbstractBroyden}
          overrides \texttt{\_compute\_tauk} $\equiv 1$.

    \item \texttt{AbstractSSBFGS} fixes $\theta_k = 0$, which simplifies the
          update to the BFGS Woodbury form.

    \item \texttt{AbstractSSDFP} fixes $\theta_k = 1$, eliminating the $v_k$
          term.
\end{enumerate}

Each abstract class has a concrete counterpart (e.g.\ \texttt{SSBroyden},
\texttt{SSBFGS}) that binds a default descent (\texttt{NewtonDescent}) and a
default search (\texttt{Zoom} line search). Users can subclass the abstract
variants to plug in alternative descents or searches. All solvers of the family
can approximate either the inverse Hessian (default) or the Hessian (option
\texttt{use\_inverse=False}). The classic \texttt{BFGS} and \texttt{DFP}, as
well as \texttt{LBFGS}, are adapted from Optimistix and also use the
\texttt{Zoom} line search by default.

Since version 0.1.0, the package uses only the public API of Optimistix and
works with Optimistix 0.0.11 and 0.1. The code described in the first arXiv
version of this note, which relied on a fork of Optimistix, is available at the
tag \texttt{arxiv-v1} of the repository, whose README lists the changes since
then.

\section{Numerical Example: PINNs for the 3D Poisson Equation}

The SSBroyden family of optimizers has recently shown improved performance over
BFGS for Physics Informed Neural Networks (PINNs) \cite{kiyani_optimizing_2025}.
In this numerical example, available as \texttt{example.py} in our repository,
we solve the Poisson equation $-\Delta u = f$ on $\Omega = [0,1]^3$ with
Dirichlet boundary conditions, where the exact solution is $u^*(x) =
    \prod_{i=1}^{3}\sin(\pi x_i)$. The solution is approximated by a fully connected
neural network (2 hidden layers of 32 units with $\tanh$ activations). The loss
function is
\[
    \mathcal{L}(\theta) = \frac{1}{2N_\Omega}\sum_{i=1}^{N_\Omega}
    \bigl(\Delta u_\theta(x_i) + f(x_i)\bigr)^2
    + \frac{1}{2N_\Gamma}\sum_{j=1}^{N_\Gamma}
    \bigl(u_\theta(x_j) - u^*(x_j)\bigr)^2,
\]
with $N_\Omega = 5000$ interior and $N_\Gamma = 800$ boundary collocation
points, sampled uniformly at random.

\begin{figure}[ht]
    \centering
    \includegraphics[width=\textwidth]{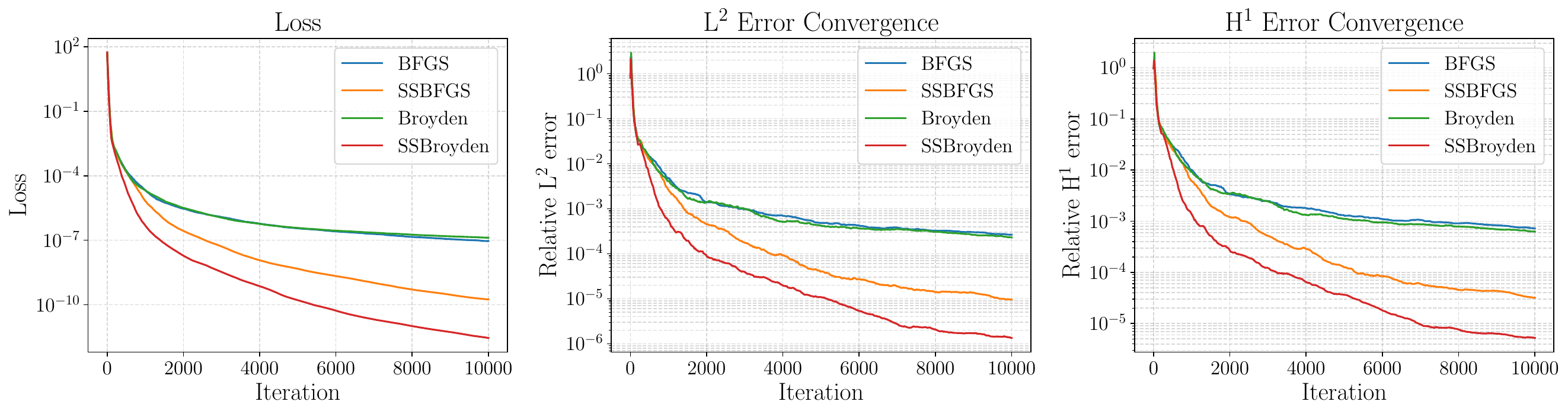}
    \caption{Convergence of quasi-Newton solvers on the 3D Poisson PINN problem.
        The self-scaled variants (SSBFGS, SSBroyden) achieve lower errors in
        fewer iterations compared to the standard BFGS and Broyden methods.}
    \label{fig:results}
\end{figure}

Figure~\ref{fig:results}, obtained with version 0.1.0 of the package (whose
code is identical to that of version 0.1.1) and its default options, compares
the convergence of the implemented solvers (BFGS, SSBFGS, Broyden, SSBroyden). The self-scaled variants converge
notably faster in terms of both loss reduction and relative $\mathrm{L}^2$ and $\mathrm{H}^1$
errors.

\section{Code Availability}
The code is available at
\url{https://github.com/IvanBioli/ssbroyden_optimistix} and on PyPI
(\texttt{pip install ssbroyden-optimistix}). All its releases, starting from
version 0.1.0, are archived on Zenodo~\cite{bioli_ssbroyden-optimistix_2026}.

\section{Use of Generative AI}
The authors used Claude Opus 5 and Claude Opus 5.5 (Anthropic) to review the
code they originally developed, to make it compatible with newer versions of
Optimistix, and to polish the text of this technical note. The authors reviewed
and edited all the resulting code and text, and take full responsibility for the
content of this note and of the accompanying software.

\section{Acknowledgements}
This work has received funding from the European Union’s Horizon Europe research
and innovation programme under the Marie Sklodowska-Curie Action
MSCA-DN-101119556 (IN-DEEP). Ivan Bioli is member of the Gruppo Nazionale
Calcolo Scientifico - Istituto Nazionale di Alta Matematica (GNCS-INdAM).

\printbibliography
\end{document}